\documentclass[12pt,preprint]{aastex}
\begin{document}

\title{A White Dwarf Merger Paradigm for \\
    Supernovae and Gamma-Ray Bursts}

\author{John Middleditch}
\affil{Modeling, Algorithms, \& Informatics, CCS-3, MS B265,
Computers \& Computational Science Division,
Los Alamos National Laboratory, Los Alamos, NM 87545}
\email{jon@lanl.gov}

\begin{abstract}
Gamma-ray bursts can appear to be a hundred times as luminous
as supernovae, but their underlying energy source(s) have
remained a mystery.  However, there has been evidence
for some time now of an association of gamma-ray bursts with
supernovae of Type Ib and Ic, a fact which has been exploited by
a number of models, to explain the gamma-ray burst phenomenon.
Here we interpret the results of basic observations of SN 1987A
and of pulsars in globular clusters, to propose the energy source,
which powers at least some long-duration gamma-ray bursts,
as core-collapse following the merger of two white dwarfs,
either as stars or stellar cores.  The beaming and intrinsic 
differences among gamma-ray bursts arise, at least in part, from 
differing amounts and composition of the gas in the merged stellar 
common envelopes, with the more energetic bursts resulting from 
mergers within less massive envelopes.  In order for the beams/jets 
associated with gamma-ray bursts to form in mergers within massive 
common envelopes (as with SN 1987A), much of the intervening 
stellar material in the polar directions must be cleared out by the 
time of core-collapse, {\it or} the beams/jets themselves must 
clear their own path.  The core-collapse produces supernovae of 
Type Ib, Ic, or II (as with SN 1987A, a SNa IIp), leaving a 
weakly magnetized neutron star remnant with a spin period near 2 
milliseconds.  There is no compelling reason to invoke any 
{\it other} model to account for gamma-ray bursts.  Far from being 
an unusual event, SN 1987A is typical, having the same merger 
source of initiation as 95\% of all supernovae, the rare exceptions 
being Ia's induced via gradual accretion from a binary companion, 
and Fe catastrophe II's.

\end{abstract}



\section{Introduction}

Gamma-ray bursts (GRBs) were discovered over three decades ago
(Klebesadel, Strong, \& Olson 1973),
with two distinct temporal classes emerging: short and long,
with distributions centered near 0.3 and 20 s, being nearly
separated at $t = 2 s$, and having larger and smaller hardness ratios
respectively \citep{pac99}, though with nearly 50\% overlap
\citep{kou93}.  That longer duration GRBs at cosmological distances
should be associated with fading afterglows was predicted a decade ago
\citep{paczyn93,kat93}, and confirmed more recently in the
optical and X-ray bands, as for GRB 970508 \citep{cos97,vpa97},
and later in the radio, as for GRB 970703 \citep{fra97}.
The identification and redshift determination of the host galaxy
for GRB 970508 definitively confirmed its extragalactic nature
\citep{met97}.  The next year Galama et al.~(1998) found the Ic SN 1998bw
within the error circle of GRB 980425 \citep{pia99}.  Just afterward
Bloom et al.~(1999) measured a possible SNa Ic component in the fading
afterglow of GRB 980326, which led to
other such possible associations, as for GRB 970228.
That GRB 980425 was four orders of
magnitude less luminous than typical cosmological bursts and the
low redshift of SN 1998bw (0.0085), led to widespread skepticism of
this association \citep{whe00}.  Recently, however, the case for a
SNa-GRB association has been bolstered by the association of
GRB 030329 with SN 2003bh \citep{hjo03}.

Supernovae are generally classified according to their spectra,
Type II for those with hydrogen lines and Type I for those
without \citep{fil97}.  These are each further divided into three
subclasses, with Ia's showing strong Si lines in their early spectra,
indicating a thermonuclear runaway of a white dwarf star, while Ib's show
helium, and Ic's neither hydrogen nor helium, but intermediate
mass elements instead.  Ib's and Ic's are thought to share core-collapse 
as a common cause with all Type II's, except that their progenitors 
have been stripped of hydrogen (Ib), or both hydrogen and helium (Ic).
Ia's also differ from Ib's, Ic's, and II's, in their luminosities, in some
cases producing, via C-O burning, nearly a solar mass of $^{56}$Ni -- ten
times the amount typical of all other types of SNe.

A number of models have been proposed to explain GRBs.  Some
earlier ones, which may not survive a GRB-SNa association,
invoke neutron star (NS) collisions with other NSs or black holes
(BH) (Lattimer \& Schramm 1974; Eichler et al.~1989; Narayan, Paczy\'nski,
\& Piran 1992).  Other, more recent models
produce SNa-like explosions by invoking collapses of
massive stars, into magnetars \citep{whe2000}, and black holes:
with strong magnetic fields via ``hypernovae'' and ``supranovae''
\citep{iwa98,vie99}, and without, via ``collapsars'' \citep{mac99}.

This paper shows that a conceptually very simple model involving
the merger of two white dwarfs (WDs), as cores of massive stars
or not (``double degenerate''), satisfies the observational constraints of
long duration GRBs and most SNe well, while others fall short of this goal.

\section{Background}

In 1987 my colleagues and I discovered the first pulsar in a
globular cluster (GC), with a period of 3 milliseconds
\citep{lyn87}.  It soon became clear that there were many more
millisecond pulsars (MSPSRs) in the GCs
than their supposed progenitors, the X-ray binaries, could
account for under the standard ``recycling'' theory \citep{alp82},
by about a factor of 100 (Chen, Middleditch, \& Ruderman 1993, hereafter
CMR).  In recycling, an ancient, solitary, slowly-spinning NS is spun up by
accretion from a captured companion.  In spite of more than 16
years of effort, recycling has not yet successfully accounted
for the GC MSPSR population.  The most careful treatment
of this issue confirmed that there had to be a
mechanism, other than (gradual) post-collision accretion,
either from companions or disruption disks, which could form
weakly magnetized, rapidly-spinning MSPSRs \citep{bai96}.

The only other way to get a WD in a GC, with a mass exceeding the
Chandrasekhar limit of 1.4 solar, is to merge two (or more) WDs
\citep{chl93}.  If core-collapse via merger is possible, then
formation of MSPSRs via merger, particularly with {\it each} having a 
binary companion, almost always dominates recycling.
This is due in part because the cross section for binary-binary 
collisions is always larger than those of the binary-single, and, 
{\it a fortiori}, the single-single collisions, necessary for the 
isolated NS star to capture a companion from which to eventually 
accrete matter.  Since the number density of binaries is approximately 
equal to that of isolated stars, the inequality of cross sections
applies to collision rates as well.  Moreover, each of the 
{\it two} stars left over from the binary-binary merger process could 
persist as post-SNa binary companions, consistent with the high 
incidence of binarity among MSPSRs.  The recycling yield for 
MSPSRs may also be low, relative to the merger yield, because evolution, 
accretion, and spinup must follow the capture, as opposed to 
just core-collapse following merger.  

The same year also saw the SN 1987A outburst, followed shortly by
the discovery of the ``mystery spot'' (Meikle, Matcher,
\& Morgan 1987; Nisenson et al.~1987).
There is now evidence for two spots \citep{nis99} on opposite sides of,
and in line with, the axisymmetric ejecta \citep{wan02}.
The closest spot was $\sim$0.06 arc s south of SN 1987A (17 light
days in projection), and had a luminosity nearly 5\% of maximum light
(3$\times$10$^{42}$ ergs/s, 8$\times$10$^8$ L$_{\bigodot}$, or 
magnitude 5.7 vs 2.5 at 6585 $\rm{\AA}$). Like the overabundance of MSPSRs 
in the GCs, this feature has never been reconciled with traditional 
models (of SNe).

\section{The Paradigm for SN 1987A}

When the structure and kinematics of the inner ring surrounding
SN 1987A became clear almost a decade ago
\citep{jak91,pla95,bur95}, other colleagues proposed a binary
merger scenario \citep{che94}.  In this picture, the
inner ring was formed by mass loss through at least one of the two
outer mass axis Lagrangean points (L2 \& L3), efficiently producing its
extremely low 10 km/s expansion velocity (as compared to blue supergiant 
(BSG) winds), essentially the thermal velocity of hydrogen at photospheric
temperatures \citep{lub76}.  The polar gradients of the potential
just beyond L2 and L3, may have helped collimate the gas outflow to
the observed small inner ring angular height.  Neither member of the close binary
was likely to have had a recent red supergiant wind
due to limitations of space.

The two fainter, outer rings around SN 1987A were formed close to
the epoch of the contact binary during which the inner ring was formed,
as all three are expanding homologously \citep{cro00}.  Radiation pressure
from one or both stars on their companion's outer atmosphere may have
produced the wind of $\sim$25 km/s, least disturbed by the orbital motion
for the two directions, nearest to, but at least 30 degrees from, their respective
poles (Chanan, Middleditch, \& Nelson 1976).  The actual half angle of each cone, formed
by the outer rings with the stellar remnant, which perhaps could be used to determine
the pre-merger mass ratio, is a more realistic 48 degrees, with the flow,
over the whole domain of polar angles, forming the observed fans of extended
emission \citep{bur95}.  Finally, the lower nitrogen abundance, relative
to that of the inner ring \citep{pan96}, indicates a more superficial
(and/or selective) source than Roche lobe overflow, for the gas in the
outer rings, consistent with the wind hypothesis.

\section{The Paradigm for the General WD-WD merger}

As the WD cores move closer to each other in their decaying orbits 
within the common envelope, the shape of that envelope must 
transition from bi-lobed to spheroidal.  Thus, for a brief 
period, the nearly merged star was likely to have been concave 
in both polar directions, but still later convex.  The transition 
between the two geometries may cause polar ejection of material, 
that we can speculate formed the target into which a polar beam/jet 
dumped (an isotropic) $10^{49}$ ergs {\it en passant} (Meikle et al.
1987).  To be conservative, the spot is assumed {\it not} to be material 
traveling at relativistic speed, prior to being hit by the beam/jet, 
which {\it was} relativistic \citep{nis99}.
 
Just prior to core-collapse, angular momentum transfer from the two
merging WD cores might have at least partially cleared out the star's
inner polar regions.  Having the SNa-associated beam/jet itself blast
through, or carry along, the normally intervening stellar material, 
seems improbable. The pre-SNa clearing process may expose the 
hotter interior of the star to the outside world.  If this exposure 
is sufficiently abrupt, the increase of apparent mean temperature and 
luminosity of the star could be exploited to form an ``early warning system'' 
for merger SNe in most of the BSGs in the Local Group and blue stragglers in 
the globulars.  It must, however, be followed promptly by core-collapse so 
that nearby material doesn't close in again.

The merged WD would rotate with a period near 1.98 s, set by the
branching between Jakobi and Maclaurin configurations \citep{che94},
and, if its mass exceeded 1.4 solar, core-collapse would follow in
many \citep{sai98} or most cases.  A NS/pulsar with a spin
period near 2 ms would form within a Type Ib, Ic, or II SNa, consistent,
in the case of the IIp SN 1987A, with the 2.14 ms signal
(Middleditch et al. 2000, hereafter M2000).

\section{Discussion}

It has already been suggested that the mystery spot in SN 1987A, about
24 light days distant from the pre-explosion binary's pole, was a result
of a lateral GRB \citep{cen99}.  But the crucial link to understanding GRBs 
may be that most Type Ib, Ic, \& II SNe, which may comprise more than 90\%
of all SNe, are the result of WD-WD mergers. Since, as we have argued,
SN 1987A was the result of such a merger, and certainly produced a NS
\citep{bio87,hir87}, then we know that such mergers can indeed produce
NS remnants, which can also be MSPSRs (M2000).  The vast overabundance, 
and other details of the MSPSRs in the GCs (CMR), including the 2.1 ms 
minimum period \citep{cam00}, strongly support this assertion.

The recent discovery of five transient X-ray MSPSRs (TXRMSPs) in the 
Galactic plane (Wijnands 2003) does not challenge this view.  The 
same mechanism, which likely produces most of the binary and solitary 
MSPSRs in the globulars, works at least as well as recycling (see above), 
even in the Galactic plane.  The discovery of more TXRMSPs in the Plane, 
without any in the GCs, may indicate that the companions, whose evolution 
drives the transient accretion, are not common among the offspring companions 
to the MSPSRs produced by mergers in the GCs.  The magnetic fields
of the NSs associated with TXRMSPs \citep{psa99} may also be larger than 
the 10$^{8-9}$ G typical of the MSPSR fields in the GCs. Either way, 
the situation is becoming embarrassing to the application of recycling 
everywhere, all the while not even exclusively supporting recycling in 
the Plane.

Core-merger for SNe engenders a natural beam/jet collimation mechanism 
and, in doing so, produces an axisymmetric SNa, consistent with SN 1987A 
\citep{wan02}, and the significant polarizations observed in SNe 
\citep{leo00,wan01}. It also doesn't suffer from the difficulty that 
plagues the massive star-black hole models, in producing the apparent 
maximum energy observed in the radio lobes \citep{fra01}.
The less massive common envelopes in such mergers may 
result in more energetic GRBs and less highly collimated beam/jets, as 
they are both easier to break through, and to accelerate in part, hence 
the association with SNe Ib's and Ic's.  However, it is not yet clear how 
much collimation is due to the merging cores or to the stellar interior 
external to them.  The GRB rate of $1.8 \times 10^{-10}$/yr/Mpc$^3$ 
\citep{sch99}, and the SN Ib,c rate of $3 \times 10^{-5}$/yr/Mpc$^3$, 
can be reconciled if each pole of the merger has a beam/jet collimated 
to a 0.5 degree diameter (Lamb, Donaghy, \& Graziani 2003b).  
The metallicity of the common envelope may also affect the GRB's.

Unlike most Type Ib and Ic SNe, the beam/jet from SN 1987A entrained 
and/or encountered a much higher density of material, and thus may have
been attenuated so that radio lobe formation did not occur (as these were 
not detected), or what radio emission was generated may have been 
absorbed or beamed.  However, simple plasma absorption for radio 
frequencies from 1 to 40 GHz would require a free-electron density of 
$\sim$10$^{10-13}$ cm$^{-3}$, but this may to be too high to achieve 
(the number density of the inner ring is $\sim$10$^4$ cm$^{-3}$).  
Synchroton self-absorption also may not occur, though the issue is 
highly complex.  

Beaming, in general, removes the $10^{54}$ erg energy requirement 
for GRBs that hypernovae, supranovae, collapsars, and other models 
were proposed to satisfy \citep{iwa98,vie99,mac99}.  
The vast majority of Type Ib 
and Ic SNe, which show no evidence of the central engine as was 
associated with SN1998bw (Berger et al. 2003), near GRB980425 \citep{kul98}, 
may also have had more massive common envelopes, and/or radio lobes 
hidden by beaming, with the additional caveat that perhaps not all 
Type Ib and Ic SNe result in a NS remnant. The yield can not be so 
low, however, as to deprive the GCs of their quota of MSPSRs.

The extremely high expansion velocities seen in SNe associated with GRBs 
(Iwamoto et al.~1998), may be due to less massive/lower metallicity common 
envelopes.  Likewise, the suggested correlation between the isotropic GRB 
energy with distance \citep{lam03}, if real, could also be a due to lower 
metallicity in the earlier common envelopes.  Although the magnetic fields 
resulting from merger are typically low (few$\times 10^9$ G -- M2000), 
these will likely have some effect on the beam/jet.  Additional energetic 
events following the initial burst could simply be due to the beam/jet 
hitting other gas targets farther away in the polar directions of the binary 
merger.  GRBs which are {\it preceded} by SNe, as might be the case for GRB 
021211 \citep{del03}, could have been produced by a beam/jet, which was 
slightly off our line of sight, hitting such a target.

Finally, magnetars, NSs with magnetic fields as high as $10^{16}$ gauss, 
which are supposed to drive the GRB/SNa explosion \citep{whe2000}, are 
observationally constrained during the SNa decline \citep{mid84}, and 
later (as in Cas A -- Chakrabarty et al.~2001) by the expected X-ray 
modulation of the still hot NS, via residual accretion or magnetic-thermal
interaction.  These would otherwise be an observer's dream, shining
through the opacity from time to time, all the while slowing down
dramatically.  The nearby reality, however, is far different, with
SNe fading time after time, like SN 1987A, at least as rapidly as
the $^{56}$Co decay curve for the first few years, with rare exceptions
likely only due to circumstellar material.  The magnetar hypothesis
can only be made viable, for the vast bulk of SNe, by the
most arduous fine tuning of spin frequency and remnant opacity
history.

The issue of whether most SNe Ia also result from merger is beyond
the scope of this work.  There could be enough such mergers, with
sufficient total mass, to account for all SNe Ia (Liebert, Arnett, \&
Benz 1997), unless most of these result in core-collapse.  Thermonuclear
runaway, however, could produce overluminous SNe Ic with the
usual lines from intermediate mass elements -- objects which have
never been observed.  Thus, unless these {\it never} happen, which
seems unlikely, such mergers {\it always} produce core-collapse,
{\it or} by the time of the thermonuclear runaway most of the
intermediate mass elements have already been consumed.  There is no
constraint on the delay between merger and runaway.  In spite of the
recent detection of hydrogen lines in the field of SN 2002ic
\citep{ham03}, the science of SNe Ia's remains in the Dark Ages.

\section{Conclusion}

We have shown that a very simple model for binary WD-WD merger
explains the details of the axisymmetric remnant expanding from, and
the ringed gas structure surrounding, SN 1987A \citep{che94}, as well as
the apparent 2.14 ms pulsar signal for the compact remnant (M2000).
By extending this picture to other Type II, as well as Type Ib and Ic
supernovae, which comprise at least 90\% of all SNe, we have accounted
for the anisotropies recently discovered in many other SNe, as well as
all of the fundamental details of long-duration GRBs.
At this point, other models to explain this class
of GRBs appear to be superfluous.

The nature of the energy source behind GRBs can {\it definitively} be
confirmed only, perhaps, for pulsars, from magnetars to MSPSRs,
by detecting an underlying periodicity. Thus, taking ``imposter'' SNe
into account, high time-resolution observations of {\it all} nearby ($< 50$ Mpc)
SNe should be made, using the largest and most sensitive optical and radio
telescopes, to search for pulsar remnants. Unlike SN 1987A, opacity will not
be an issue for many of these.  Observations of only SNe which happen to be
associated with GRBs might not be fruitful, as these are usually much
farther away, and the proximity of the line of sight to the rotation
axis of the binary merger could reduce the amplitude of the pulsar signal.

Although a number of extragalactic SNe (but no Ib's or Ic's) have already 
been searched \citep{mid84}, and this has continued sporadically, without 
success, only scant resources have been devoted to this effort so far. 
Continuing indefinitely {\it without} making further, more sensitive searches, is 
to risk wasting time and resources.  The results from SN 1987A, now
less controversial due to strong evidence for precession in two other
``normal'' pulsars, B1828-11 and B1642-03 (Stairs, Lyne, \& Shemar 2000; 
Shabanova, Lyne, \& Urama 2001), imply a pulsed optical 
signal of 10 or more solar luminosities, at 5.0 -- 6.5 years of age (M2000).  
This is hardly surprising, as the Crab pulsar's (nearly entirely pulsed) 
optical output is 4 solar, and nanosecond radio bursts in pulsars are bright 
enough to be seen in neighboring galaxies \citep{han03}.

This new understanding of GRBs and SNe may represent the end of an era, 
begun four decades ago, of exploration and discovery in high-energy 
astrophysics.  The core-collapse events, and early stages of the 2 ms 
pulsar remnants from SNe in the Local Group, may be detectable by 
gravitational wave detectors such as LIGO (Santostasi, Johnson, \& Frank 2003), 
provided an early, rudimentary spin frequency history can be obtained by 
detecting the pulsars in other bands, such as the optical and/or radio.  
Thus this new understanding also {\it begins} an era, which will span the 
first part of our own 21$^{st}$ Century, of extragalactic pulsar astronomy.

\acknowledgements

I thank Kaiyou Chen, Stirling Colgate, and Chris Fryer for
their support, guidance, conversations and  insight, and
Alexander Heger for timely comments.  This work  was performed
under the auspices of the Department of Energy.

\end{document}